\documentclass[11pt]{article}
\usepackage{moriond,epsfig}
\usepackage[latin1]{inputenc}
\usepackage[OT1]{fontenc}

\usepackage{graphicx}

\begin{document}
\vspace*{4cm}
\title{The VIMOS-VLT Deep Survey: The dependence of clustering on galaxy stellar mass
  at $z\sim 1$}

\author{B. Meneux}
\address{Max Planck Institut f\"ur extraterrestrische Physik, D-85741, Garching, Germany\\
  Universit\"ats-Sternwarte M\"unchen, Scheinerstrasse 1, D-81679 Munich, Germany}
\author{L. Guzzo}
\address{INAF-Osservatorio Astronomico di Brera - Via Bianchi 46, I-23807 Merate (LC),
  Italy }

\author{and the VVDS collaboration\footnote{B. Garilli,
O. Le F\`evre,
A. Pollo,
J. Blaizot,
G. De Lucia,
M. Bolzonella,
F. Lamareille,
L. Pozzetti,
A. Cappi,
A. Iovino,
C. Marinoni,
H.J. McCracken,
S. de la Torre,
D. Bottini,
V. Le Brun,
D. Maccagni,
J.P. Picat,
R. Scaramella,
M. Scodeggio,
L. Tresse,
G. Vettolani,
A. Zanichelli,
U. Abbas,
C. Adami,
S. Arnouts,
S. Bardelli,
A. Bongiorno,
S. Charlot,
P. Ciliegi,
T. Contini,
O. Cucciati,
S. Foucaud,
P. Franzetti,
I. Gavignaud,
O. Ilbert,
B. Marano,
A. Mazure,
R. Merighi,
S. Paltani,
R. Pell\`o,
M. Radovich,
D. Vergani,
G. Zamorani,
E. Zucca
}}
\address{}

\maketitle
\abstracts{
We have investigated the dependence of galaxy clustering on their stellar mass at
z$\sim$1,
using the data from the VIMOS-VLT Deep Survey (VVDS).
We have measured the projected two-point correlation function of galaxies, $w_p(r_p)$
for a set of stellar mass selected samples at an effective redshift $<z>=0.85$.
We have control and quantify all effects on galaxy clustering due to the incompleteness
of our low mass samples.
We find that more massive galaxies are more clustered. When compared to similar results
at $z\sim0.1$ in the SDSS, we observed no evolution of the projected correlation function
for massive galaxies. These objects present a stronger linear bias at $z\sim1$ with
respect to low mass galaxies. As expected, massive objects at high redshift are found
in the highest pics of the dark matter density field.
}

\section{Introduction}

In the currently accepted scenario, galaxies are thought to form within
extended dark-matter halos\cite{whiterees1978}, which grow through
subsequent mergers in a hierarchical fashion. A major challenge in testing
this general picture is to connect the observable properties of galaxies to
those of the dark-matter halos in which they are embedded, as predicted, e.g.,
by large n-body simulations\cite{springel2006}. 
Recent theoretical works seem to indicate that a fairly direct relationship
indeed exists between global galaxy properties (e.g.~their stellar or 
total baryonic mass, or their luminosity) and the halo mass, {\it before} it
is accreted by a larger dark-matter halo\cite{conroy2006,wang2006,wang2007}.

At the current epoch and at high redshift, the observed clustering of galaxies
is found to depend significantly on their specific properties, such as
luminosity\cite{guzzo2000,norberg2002,pollo2006,coil2006}, 
color or spectral type\cite{zehavi2005,meneux2006},
morphology\cite{giovanelli1986,guzzo1997}
and stellar mass\cite{li2006}.
The dependence of clustering on stellar
mass in the local Universe has been shown in the Local Universe, from a large sample of
galaxies of the Sloan Digital Sky Survey\cite{li2006}. Galaxies of larger mass are seen
to be more clustered than low-mass ones, with the effect increasing above the
characteristic knee value $M^*$ of the Schechter mass function.
Using the VIMOS-VLT Deep Survey (VVDS), we have investigated 
for the first time the evolution with redshift of the galaxy clustering as a function
of stellar mass at $z\sim1$\cite{meneux2008}.


The 
VVDS is performed with the VIMOS multi-object
spectrograph at the ESO Very Large Telescope\cite{lefevre2003_spie} and
complemented with multi-color 
imaging data obtained at the CFHT and NTT
telescopes\cite{hjmcc2003,lefevre2004_im,iovino2005,radovich2003}.
We have used the so-called ``first epoch'' data, collected in the VVDS-02h
``Deep'' field.  This is a purely magnitude limited sample to
$I_{AB}=24$, covering an area of $0.49$ square degrees with a mean sampling of
$\sim23\%$. The full sample includes $\sim11000$ redshifts.
Details about observations, data reduction, redshift measurement and
quality assessment can be found elsewhere\cite{lefevre2005_vvds}.

We use a Concordance Cosmology with $\Omega_m = 0.3$ and
$\Omega_{\Lambda} = 0.7$. The Hubble constant is normally parametrised via
$h=H_0/100$ to ease comparison with previous works. Stellar masses are quoted
in unit of $h=1$. All correlation length values are quoted in comoving
coordinates. 

\section{Stellar mass and incompleteness}

Stellar masses for all galaxies in the VVDS catalogue were estimated
by fitting their Spectral Energy Distribution, as sampled by the
VVDS multi-band photometry, with libraries of stellar population
models\cite{bc2003}.
The robustness of the method and intrinsic errors have been fully
tested\cite{pozzetti2007}.
The typical error on stellar mass is $\sim$O.1~dex.

We concentrated our study of the dependence of galaxy clustering on stellar mass
in the redshift range [0.5-1.2].
Such a sample includes 3218 galaxies more massive than $10^9~M_\odot$.
However, the VVDS is a flux limited survey, and even if the luminosity is correlated
to the stellar mass,
this relation has some scatter and then, some faint galaxies might be missed in a given
stellar mass range.
We investigated the completeness limit using the data themselves and 40 mock catalogues
complete in stellar mass
built from the Millennium run dark-matter outputs\cite{springel2005} coupled to semi-analytical
models\cite{delucia_blaizot2007}.

\begin{figure}[h]
  \begin{center}
    \includegraphics[width=8.5cm]{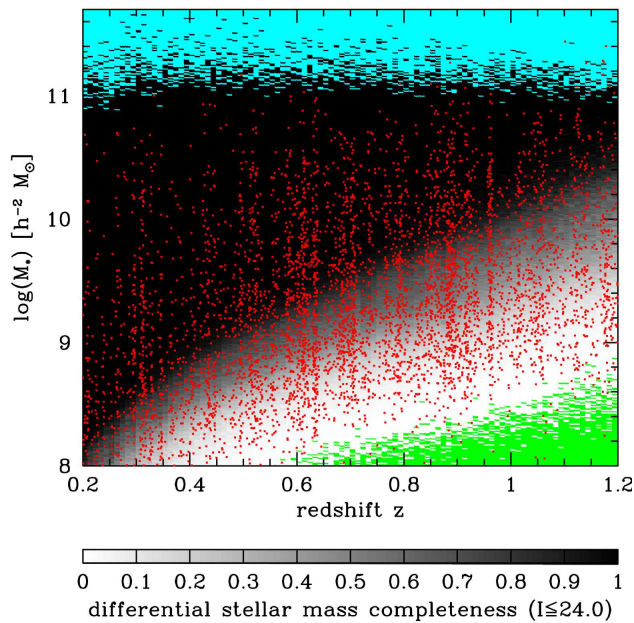}
  \end{center}
  \caption{Stellar mass completeness at $I_{AB}=24$ as a function of redshift estimated
    from 40 mock catalogues. 
    The green area indicates regions where all galaxies are fainter than the apparent
    luminosity thresold.
    The blue area shows region where no galaxy was found in the mock at any $I_{AB}$
    apparent magnitude.
    The red points correspond to the VVDS data.
  }
  \label{fig:completeness}
\end{figure}

As an illustration, Fig.~\ref{fig:completeness} shows the stellar mass vs. redshift relation
of the mock catalogues, with, coded in grey scales,
the fraction of galaxies brighter than $I_{AB}=24$ in narrow redshift ($\Delta z=0.01$)
and stellar mass ($\Delta \log(M)=0.01$) bins. 
According to these numerical simulations, we miss $\sim30$\% of galaxies in the stellar
mass range $\log(M/M_\odot)=[9.5-10]$ between redshift z=0.5 and z=1.2.
Samples of galaxies more massive than $10^{10}~M_\odot$ are almost complete.
It is interesting to note that, rather realistically, the typical galaxies producing
the incompleteness at faint fluxes in the mock samples are red objects with high
mass-to-light ratio.

\section{Galaxy clustering as a function of stellar mass}

We measure the two-point correlation function $\xi(r_p,\pi)$ and
its projection along the line of sight $w_p(r_p)$\cite{dp83}
using standard estimator\cite{lansal1993}. We take into account the complex observing
strategy of the survey.
We checked using the mock catalogues the effect of stellar mass incompleteness on the
clustering measurement, comparing the projected correlation function for mock samples
complete in stellar mass and selected at $I_{AB}\le 24$.
We concluded that, when incompleteness is the strongest
(i.e.~for $\log(M/M_\odot)=[9-9.5]$),
the amplitude of the correlation function is underestimated on small scales by a factor
up to 2\cite{meneux2008}.

Taking into account these effects in our conclusions, we measured the projected
correlation function for several stellar mass samples from the real data at $z\sim1$.
We observed for the first time at this redshift that more massive galaxies are more
clustered than low mass galaxies, the correlation length $r_0$ increasing from
$2.76_{-0.15}^{+0.17}$~h$^{-1}$Mpc for $\log(M/M_\odot)\ge 9$ to
$4.28_{-0.45}^{+0.43}$~h$^{-1}$Mpc for $\log(M/M_\odot)\ge 10.5$.
This dependence on stellar mass was predicted by hydrodynamical
simulations\cite{weinberg2004} and is also
observed in the mock catalogues we have been using\cite{meneux2008}.

\begin{figure}[h]
  \includegraphics[width=\textwidth]{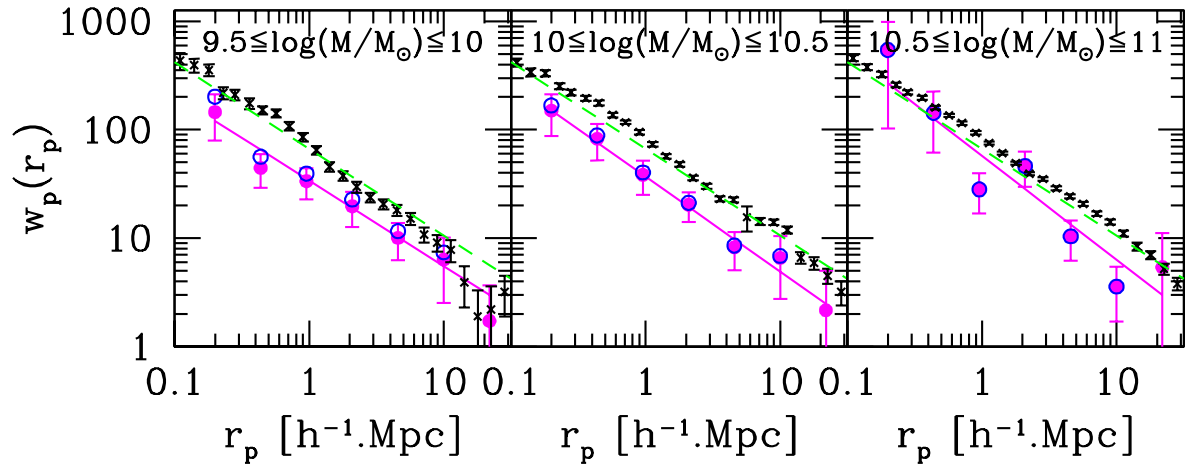}
  \caption{Comparison of the projected correlation function $w_p(r_p)$
    from SDSS data 
    at z$\sim$0.15 ({\it black cross}) to VVDS measurement at z$\sim$0.85
    ({\it magenta filled dots and solid lines})
    in 3 ranges of stellar masses.
    The green dashed line is a power-law reference line drawn with
    $r_0=5$~h$^{-1}$.Mpc and $\gamma=1.8$. Error bars on VVDS
    measurement have been estimated from the variance among 40 mock
    catalogues. Blue open circles indicate VVDS measurements corrected
    for stellar mass incompleteness.
  }
  \label{fig:wp_vvds_sdss}
\end{figure}

We then compared our measurement to similar analysis made at $z\sim0.1$ with SDSS
data\cite{li2006}.
Figure~\ref{fig:wp_vvds_sdss} shows a faster evolution of the correlation function
$w_p(r_p)$ for galaxy less massive than $10^{10.5}~M_\odot$, while the clustering properties
of massive galaxies do not appear to evolve with time.

Thereby, the most massive galaxies display an evolution of their linear bias factor
(from $b_L=1.62\pm 0.18$ at z$\sim$1 to $b_L=1.40\pm 0.03$ at z$\sim$0.1),
while it remains roughly constant for lower mass objects ($b_L\sim1.3$).
Massives galaxies are found at $z\sim1$ in the highest peaks of the density
field\cite{meneux2008}.
This finding is expected in a hierarchical scenario in which the
most massive peaks of the mass density field collapse earlier and
evolve faster\cite{mowhite1996}.
This interpretation qualitatively supports a
scenario in which the stellar mass of a galaxy is essentially proportional
to the mass of the dark-matter halo in which it was last the central
object, consistent with recent simulations\cite{conroy2006,wang2006}.

\section{Acknowledgments}

BM thanks the organisers of the conference for the financial support.
LG gratefully acknowledges the hospitality of MPE, MPA and ESO.

This research program has been developed within the framework of the VVDS
consortium and has been partially supported by the CNRS-INSU and its
Programme National de Cosmologie (France), and by Italian Ministry (MIUR)
grants COFIN2000 (MM02037133) and COFIN2003 (num.2003020150).
The VLT-VIMOS observations have been carried out on guaranteed time
(GTO) allocated by the European Southern Observatory (ESO) to the
VIRMOS consortium, under a contractual agreement between the Centre
National de la Recherche Scientifique of France, heading a
consortium of French and Italian institutes, and ESO, to design,
manufacture and test the VIMOS instrument.

\end{document}